\documentclass[prb,twocolumn,superscriptaddress,showpacs,amsmath]{revtex4}

\usepackage{graphicx}
\usepackage{bm}

\begin{document}

\title{Massive Dirac surface states in topological insulator/magnetic
  insulator heterostructures}
\author{Weidong Luo}
\affiliation{Geballe Laboratory for Advanced Materials, Stanford
  University, Stanford, California 94305, USA}
\author{Xiao-Liang Qi}
\affiliation{Department of Physics, Stanford University, Stanford,
  California 94305, USA}

\date{\today}

\begin{abstract}
Topological insulators are new states of matter with a bulk gap and
robust gapless surface states protected by time-reversal
symmetry. When time-reversal symmetry is broken, the surface states
are gapped, which induces a topological response of the system to
electromagnetic field--the topological magnetoelectric effect. In this
paper we study the behavior of topological surface states in
heterostructures formed by a topological insulator and a magnetic
insulator. Several magnetic insulators with compatible magnetic
structure and relatively good lattice matching with topological
insulators ${\rm Bi_2Se_3}, {\rm Bi_2Se_3}, {\rm Sb_2Te_3}$ are
identified, and the best candidate material is found to be MnSe, an
anti-ferromagnetic insulator. We perform first-principles calculation
in ${\rm Bi_2Se_3/MnSe}$ superlattices and obtain the surface state
bandstructure. The magnetic exchange coupling with MnSe induces a gap
of $\sim$54 meV at the surface states. In addition we tune the
distance between Mn ions and TI surface to study the distance
dependence of the exchange coupling.
\end{abstract}


\maketitle

Topological insulators (TI) are new states of quantum matter which has
the same symmetry as the conventional insulators and semiconductors
but cannot be adiabatically deformed to them without going through a
phase transition.  Recently, time-reversal invariant (TRI) TI's are
theoretically predicted and experimentally realized in both two and
three dimensions (2D and 3D).\cite{moore2010,hasan2010,qi2011rmp} A
TRI TI is characterized by robust surface states and unique, quantized
response properties, just like the quantized Hall conductance in 2D
quantum Hall states. For 3D TI the topological response is the
topological magnetoelectric (TME) effect\cite{qi2008b}, which is a
magneto-electric effect with magnetization ${\bf M}$ generated by
electric field ${\bf E}$ with a quantized coefficient. The TME effect
occurs when the surface states of TI become gapped due to
time-reversal symmetry breaking, and is a generic property of 3D
topological insulators, which can be obtained theoretically from
generic models and from an effective field theory
approach\cite{qi2008b,essin2009}, independently of microscopic
details. Various consequences of the TME effect have been proposed,
including Faraday/Kerr rotation of linear polarized
light\cite{karch2009,maciejko2010,qi2008b,tse2010}, the image monopole
effect\cite{qi2009}, the charge carried by a mangnetic
monopole\cite{franz2010b,franz2010c},and other types of coupling
between the charge and spin degree of freedom at the TI
surface\cite{yokoyama2010b,nomura2010}. Experimental progress has been
made recently on the Faraday/Kerr effect in 3D
TI\cite{aguilar2011,laforge2010,sushkov2010,jenkins2010}, but the
quantized effect predicted have not been observed yet.

To realize the TME effect it is essential to introduce time-reversal
symmetry breaking (T-breaking) at the surface of TI to make the
surface insulating. There are two possible physical ways to open the
T-breaking gap at the surface. The first approach is to introduce
magnetic impurities such as Mn or Fe to topological insulators. Both
the Dirac-type surface states\cite{liu2009} and the bulk
states\cite{yu2010} can mediate ferromagnetic coupling between
magnetic impurities and thus induce ferromagnetic order under proper
conditions. The surface state gap induced by doping magnetic
impurities Mn and Fe has been observed in various experiments
including transport, angle-resolved photo-emission (ARPES) and X-ray
magnetic circular dichroism (XMCD)\cite{chen2010b,hasan2012,liu2012}
This approach has the advantage of simple experimental setting which
can be realized in both bulk materials and thin films. However, the
surface state gap is usually small and probably non-uniform due to the
low density and disorder effect of the impurities. This may explain
why Hall measurements on thin films have observed a large anomalous
Hall effect but the longitudinal conductivity is still
nonzero.\cite{chang2011} The second approach is to make a
heterostructure between a topological insulator and a magnetic
insulator (MI), so that the surface states are gapped due to exchange
coupling with the MI. Compared to the first approach, this approach
has the potential to achieve a stronger and more uniform exchange
coupling and thus realize insulating surface states and TME effect in
a higher temperature. The main challenge of this approach is to find
the suitable material for the MI, which can form a high-quality
heterostructure with topological insulators and lead to a strong
exchange coupling. This is the goal of the current work.

In this paper we study heterostructures of the ${\rm Bi_2Se_3}$ family
of TI with MI's. We first carried an extensive material search and
identify several materials which have relatively the best lattice
matching with TI. The candidate materials are summarized in Table
\ref{table-1} and their properties will be discussed in more detail in
the next section. It is important to notice that some
anti-ferromagnetic insulators can also be used to introduce the
surface state gap, although a ferromagnetic exchange coupling at the
surface is required. Since the exchange coupling is very short-ranged,
it will be dominated by the first layer of magnetic ions at the
interface of TI and MI. Therefore if an anti-ferromagnetic insulator
has magnetic moments which are aligned to each other in each layer
parallel to the surface, but staggering in the perpendicular
direction, it provides the same form of ferromagnetic exchange
coupling as a ferromagnetic insulator does. In fact, the best
candidate material found in our search is an anti-ferromagnetic
insulator, MnSe, which is a better candidate than ferromagnetic
insulators such as EuS since the latter has magnetism from
$f$-electrons and has a weaker exchange coupling with the
$p$-electrons in TI compared with the $d$-electrons in MnSe. We
present {\it ab initio} calculation in MnSe/${\rm Bi_2Se_3}$
superlattices, from which we obtain the surface state gap and also
describe it in a surface state effective model. In addition we tune
the distance between Mn ions and TI surface to study the distance
dependence of the exchange coupling. Then we conclude by discussing
the band-bending effect caused by the interface charge at the TI/MI
junction, which is the main experimental challenge that needs to be
addressed in future works.

\noindent{\bf Criteria and candidate materials}

The topological insulators Bi$_2$Se$_3$, Bi$_2$Te$_3$, and
Sb$_2$Te$_3$ have layered structure, and the 2-dimensional lattice
within each layer has triangular symmetry. The 2D lattice constants of
Bi$_2$Se$_3$, Bi$_2$Te$_3$, and Sb$_2$Te$_3$ are 4.1355
\AA~\cite{Bi2Se3_struc}, 4.395 \AA~\cite{Bi2Te3_struc}, and 4.264
\AA~\cite{Sb2Te3_struc}, respectively. We look for candidate MI
materials with 2D crystal plane of compatible symmetry to TI layers,
ie. hexagonal lattice. The other criteria for the candidate MI
materials include similar lattice constant, and ferromagnetic moments
in the 2D hexagonal interface atomic plane.

A list of candidate MI materials is shown in
table~\ref{table-1}. EuO~\cite{EuO_struc}, EuS~\cite{EuS_struc},
EuSe~\cite{EuSe_struc}, and MnSe~\cite{MnSe_struc} have the cubic
rocksalt structure, of which the atoms in the (111) plane form
triangular lattice, with compatible lattice constant to the TIs. Both
EuO and EuS are ferromagnetic insulators, which meets the requirement
of correct magnetic configuration, however the lattice constant of EuO
is too small for good lattice matching with the common 3D topological
insulators. EuSe has a complex magnetic phase diagram, and it also
becomes FM under suitable pressure range. Magnetism in EuO, EuS and
EuSe originates from the half-filled 4$f$ orbitals of the Eu$^{+2}$
ions. MnSe has the type-II (G-type) anti-ferromagnetic structure, of
which the magnetic moments in each (111) atomic planes are
ferromagnetic. MnTe~\cite{MnTe_struc} is chemically similar to MnSe,
although it has a hexagonal lattice. Mn moments in MnTe also have an
AFM configuration, formed of alternating hexagonal (0001) FM
planes. Magnetism in MnSe and MnTe comes from the high-spin $d^5$
orbitals of the Mn$^{+2}$ ions. RbMnCl$_3$~\cite{RbMnCl3_struc} is
also in hexagonal structure, with the Mn sites forming alternating
(0001) FM planes. Although its hexagonal plane has a much larger
lattice constant, it matches the $\sqrt{3}\times\sqrt{3}$
reconstruction of the in-plane lattice of Bi$_2$Se$_3$. Thus
RbMnCl$_3$ is a potential candidate material, although the magnetic
coupling would be weaker because only 1/3 of the atomic sites of the
TIs will be in contact with the Mn sites.

In the candidate materials listed in Table \ref{table-1}, we carried
{\it ab initio} calculations to several of them including MnSe/${\rm
  Bi_2Se_3}$, MnTe/${\rm Bi_2Se_3}$, MnTe/${\rm Sb_2Te_3}$, EuS/${\rm
  Bi_2Se_3}$ and EuSe/${\rm Bi_2Se_3}$. Among these heterostructures,
we find MnSe/${\rm Bi_2Se_3}$ to be the best one, with relatively
strong exchange-coupling and simple surface state band structure. In
the following we will focus on the results of MnSe/${\rm Bi_2Se_3}$
heterostructure, and present the results on MnTe/${\rm Sb_2Te_3}$ and
MnTe/${\rm Bi_2Se_3}$ in the appendix as a comparison.

\begin{table*}
\caption{A list of candidate magnetic insulators and their properties.}
\label{table-1}
\begin{ruledtabular}
\begin{tabular}{cllll}
Materials & Structure & Lattice matching & Magnetic phase & Comments
\\ \hline
EuO & rocksalt, & 5.145/$\sqrt{2}$ & FM & 4{\it f} electrons
\\
 & {\it a} = 5.145 \AA & = 3.64 \AA & & weak coupling \\ \hline
EuS & rocksalt, & 5.98/$\sqrt{2}$ & FM & 4{\it f} electrons
\\
 & {\it a} = 5.98 \AA & = 4.23 \AA & & weak coupling \\ \hline
EuSe & rocksalt, & 6.192/$\sqrt{2}$ & complex, & 4{\it f} electrons
\\
 & {\it a} = 6.192 \AA & = 4.38 \AA & FM under P & weak coupling \\ \hline
MnSe & rocksalt, & 5.464/$\sqrt{2}$ & type-II AFM, & 3{\it d} electrons
\\
 & {\it a} = 5.464 \AA & = 3.86 \AA & FM (111) planes & stronger coupling \\ \hline
MnTe & hexagonal, & 4.1497 \AA & AFM, & 3{\it d} electrons
\\
 & {\it a} = 4.1497 \AA & & FM (0001) planes & stronger coupling \\ \hline
RbMnCl$_3$ & hexagonal, & 7.16/$\sqrt{3}$ &  AFM, & only 1/3 of
\\
 & {\it a} = 7.16 \AA & = 4.13 \AA & FM (0001) planes & lattice points \\
\end{tabular}
\end{ruledtabular}
\end{table*}

\noindent{\bf Bi$_2$Se$_3$/MnSe interface}

\begin{figure}
\includegraphics*[width=36mm]{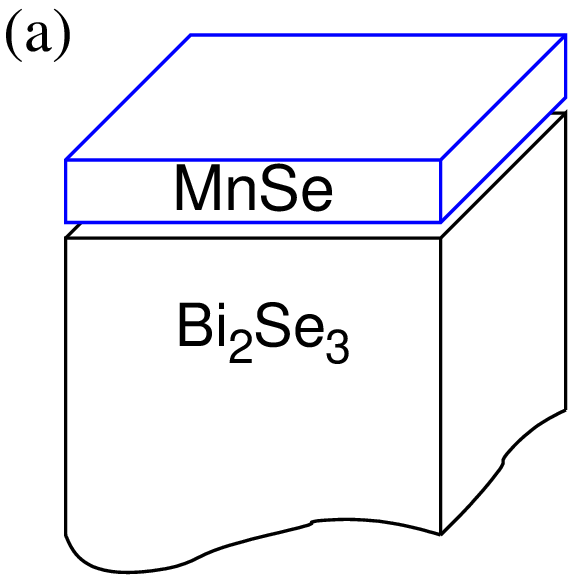}
\includegraphics*[width=30mm]{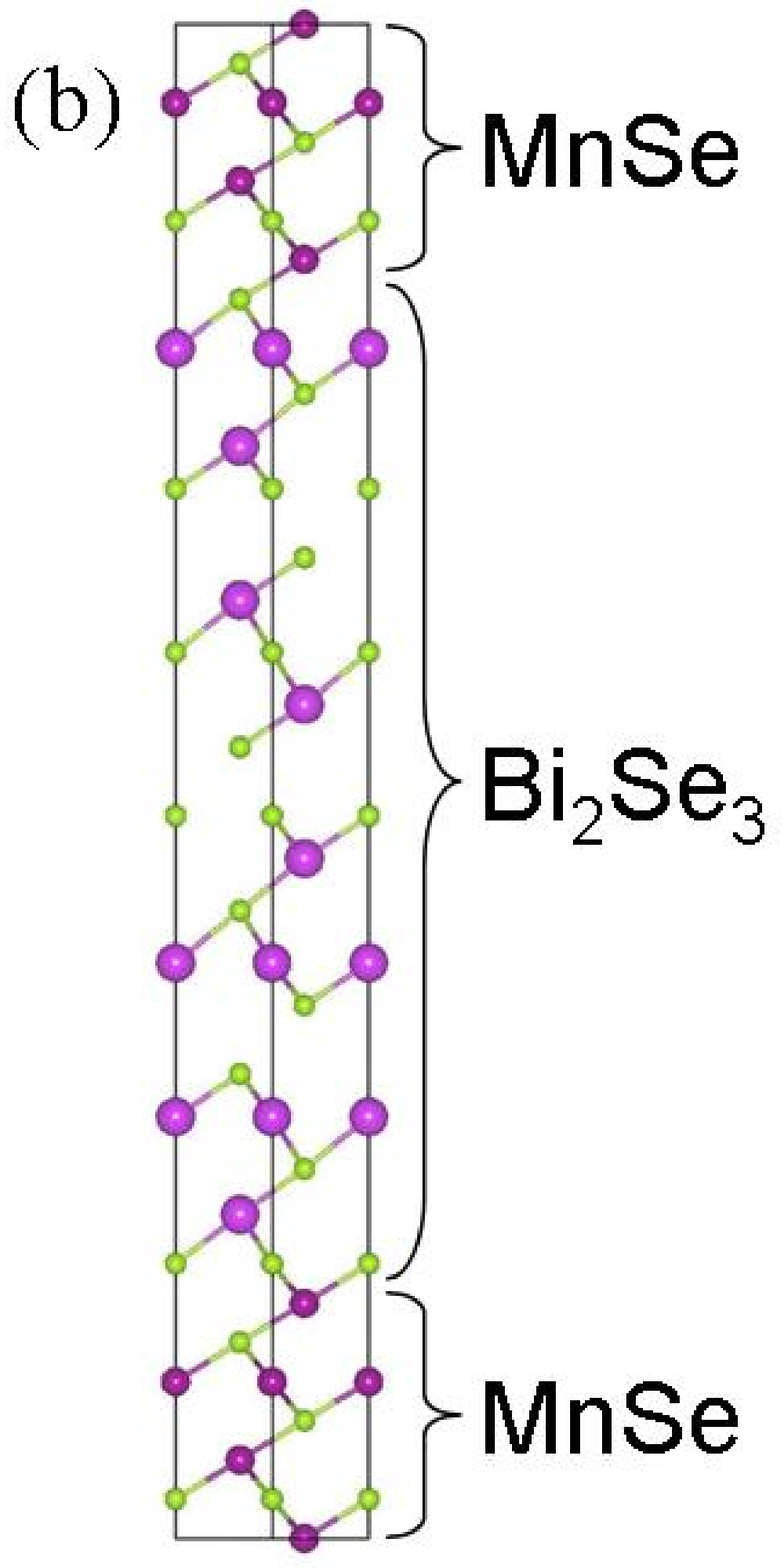}
\caption{(color online). (a) Schematic picture of a MnSe film deposited on the surface of Bi$_2$Se$_3$.
  (b) Crystal structure of the supercell structure in the first-principles calculations.}
\label{figure1}
\end{figure}

We study the Bi$_2$Se$_3$/MnSe interface by constructing a
superlattice composed of Bi$_2$Se$_3$ slab and MnSe slab. The
supercells employed in the first-principles calculations are required
to have inversion symmetry, and the Mn atoms at the top and bottom
surfaces of Bi$_2$Se$_3$ slab have parallel spin orientation. The
thickness of the Bi$_2$Se$_3$ slab is chosen to be 4 QLs, the MnSe
slab is determined to have a thickness of 6$n$+1 (7, 13, 19, $\ldots$)
layers, in order to meet the requirements of inversion symmetry and
parallel spin orientations. To simulate the physical system of a MnSe
film deposited on top of a bulk ${\rm Bi_2Se_3}$, the in-plane lattice
constant of the MnSe is fixed to that of Bi$_2$Se$_3$, and the
out-of-plane lattice constant is determined by energy
optimization. The 4-7 supercell structure (composed of 4 QLs of
Bi$_2$Se$_3$ and 7 layers of MnSe) is shown in fig.~\ref{figure1}(b).

First-principles density functional theory (DFT) calculations are
performed to optimize the supercell structure and to obtain the
electronic band structure. We used the Hohenberg-Kohn
density-functional theory with the generalized-gradient approximation
(GGA)~\cite{PBE1996} and the projector augmented-wave method as
implemented in VASP~\cite{Blochl1994,Kresse1996,Kresse1999}.The Mn
3$d$ orbitals are treated with the GGA plus Hubbard $U$ (GGA+$U$)
method~\cite{Anisimov1991, Liechtenstein1995}, and typical values of
correlation parameters are used~\cite{Youn2005,Amiri2011}: $U$ = 5.0
eV, and $J$ = 1.0 eV.

If not noted otherwise (as when tuning the separation distance between
Bi$_2$Se$_3$ and MnSe slabs), the atomic coordinates of the
superlattice are optimized as in the following. The middle part of the
Bi$_2$Se$_3$ slab is fixed to the experimental structure, while
the atomic coordinates of the whole MnSe slab together with the first Bi and
Se atomic planes of Bi$_2$Se$_3$ at the interface are optimized
according to the forces from first-principles calculations.

\begin{figure}
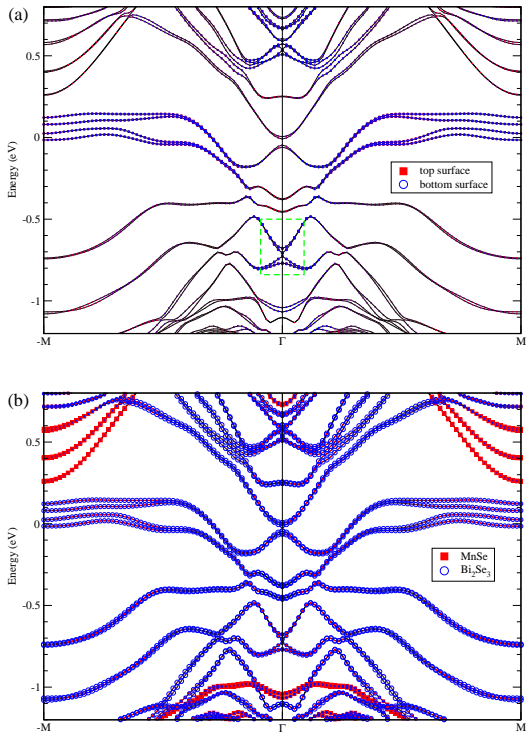

\includegraphics[width=70mm]{fig2a.eps} \\
\vspace{5mm}
\includegraphics[width=70mm]{fig2b.eps}
\caption{(color online). Bandstructure of the Bi$_2$Se$_3$/MnTe
  superlattice. The two panels show the projection of the
  wavefunctions to the Bi and Se atoms at the top and bottom surfaces
  (a), and that to the two materials Bi$_2$Se$_3$ and MnSe (b). The
  calculation is done for the superlattice structure composed of 4
  quintuple layers of Bi$_2$Se$_3$ and 13 layers of MnSe, with Mn
  spins along [001] direction.}
\label{figure2}
\end{figure}

\noindent{\it Bandstructure.--}
We first investigate the 4-13 Bi$_2$Se$_3$/MnSe superlattice, with Mn
spins along [001] direction. The bandstructure from first-principles
calculations is shown in fig.~\ref{figure2}. By projecting the bands
to the Bi and Se atoms of the top and bottom surfaces, shown in
fig.~\ref{figure2}(a), we identify the Dirac cone feature located
about 0.4 eV below the bulk gap of Bi$_2$Se$_3$ as surface states of
Bi$_2$Se$_3$. Further analysis on the spin directions of the states
around $\Gamma$ point confirms it to be the topological surface
state. A small gap appears at the Dirac point, and both the lower and
upper Dirac cones show spin related energy splitting in the vicinity
of $\Gamma$ point, indicating the magnetic interaction with MnSe.

The band gap of MnSe is much larger than that of Bi$_2$Se$_3$. In
order to obtain a clearer picture of the electronic states of the two
materials forming the superlattice, we project the bands separately to
the Bi$_2$Se$_3$ slab and the MnSe slab in the superlattice, shown in
fig.~\ref{figure2}(b). The bulk gap of Bi$_2$Se$_3$, as well as the
Dirac cone states, are located in the band gap of MnSe, which makes it
possible to realize a fully gapped system and observe the TME effect.

%


\noindent{\it Effective model fitting.--}
To gain better understanding of the exchange coupling experienced by
the surface states, we introduce an effective Hamiltonian
$H_\text{eff}$ for the Dirac fermion surface states of the topological
insulator thin film coupled with the exchange field\cite{liu2010}
\begin{widetext}
\begin{equation}
H_\text{eff} =  D k^2 I +
\begin{pmatrix}
\hbar v_F (\sigma_x k_y - \sigma_y k_x) +
\bm{M}\cdot\bm{\sigma} & t I \\
t I & -\hbar v_F (\sigma_x k_y - \sigma_y k_x) + \bm{M}\cdot\bm{\sigma}
\end{pmatrix},
\label{effective_H}
\end{equation}
\end{widetext}
Here $\sigma_i$ $(i=x,y,z)$ are the Pauli matrices, and $I$ the
identity matrix. $\bm{M}$ and $t$ correspond to the effective magnetic
field acting on the Dirac fermion surface states and the inter-edge
interaction between the two surfaces. The other parameters in the
model are: $D$, the quadratic term, and $v_F$ the Fermi velocity.

The eigen-energies of the model Hamiltonian can be solved
analytically,
\begin{equation}
E = D k^2 \pm\sqrt{k^2+M^2+t^2 \pm 2\sqrt{M^2 t^2 + (M_x k_y -M_y k_x)^2}},
\label{H_solution}
\end{equation}
here we set $\hbar v_F = 1$. We note that at $\Gamma$ point ($k=0$),
the four eigen-energies are $E = \pm (M \pm t)$. In the case of strong
magnetic coupling and weak inter-edge interaction where $M > t$, the
zone-center eigen-energies (from low energy to high energy) and the
corresponding spin orientations are: $-M-t$ (spin down), $-M+t$ (spin
down), $M-t$ (spin up), and $M+t$ (spin up). In the opposite case
where $M < t$, the zone-center eigen-energies and spin orientations
are: $-M-t$ (spin down), $M-t$ (spin up), $-M+t$ (spin down), and
$M+t$ (spin up). In both cases, an energy gap $\Delta$ forms at the
original Dirac point, $\Delta = 2 |M-t|$.

\begin{figure}
\includegraphics*[width=85mm]{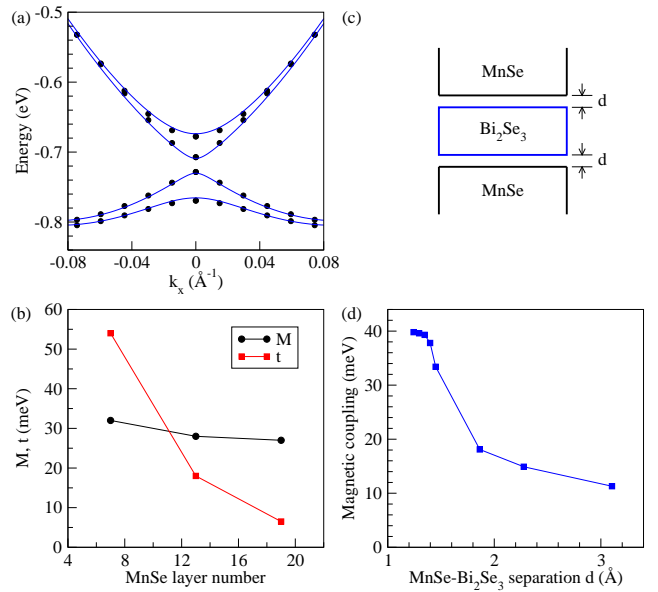}
\caption{(color online). (a) The calculated band structure (black
  circle, the same data as Fig. \ref{figure2}) is fitted by the
  effective model (\ref{effective_H}) (blue curves). (b) The
  dependence of the inter-edge interaction $t$ and the effective
  exchange field $M$ on the thickness of the MnSe slab, obtained from
  the fitting. (c) A schematic picture of the material structure used
  in the calculations shown in the next panel, with the distance $d$
  between MnSe and Bi$_2$Se$_3$ tuned (see text). (d) Dependence of
  $M$ on the separation $d$ between the Bi$_2$Se$_3$ and MnSe slabs.}
\label{figure3}
\end{figure}

The Dirac fermion bandstructure obtained from first-principles
calculations is fitted to the model solution in
equation~\ref{H_solution}. For the superlattice structure composed of
4 quintuple layers of Bi$_2$Se$_3$ and 13 layers of MnSe with Mn spins
along [001] direction, the calculated Dirac fermion surface states
around $\Gamma$ point for crystal momentum along $k_x$ ($\Gamma$-M)
direction are fitted satisfactorily to the model solution, as shown in
figure~\ref{figure3}(a). The fitting parameters are: $M$ = 28.2 meV,
$t$ = 17.6 meV, $D$ = 9.8 eV\AA$^2$, and $v_F$ = 2.66$\times$10$^5$
m/s. Fitting the model solution to the calculated band dispersion
along $k_y$ ($\Gamma$-K) direction results in very similar parameters,
$D$ = 9.9 eV\AA$^2$, and $v_F$ = 2.70$\times$10$^5$ m/s.

%


\begin{figure}
\includegraphics*[width=85mm]{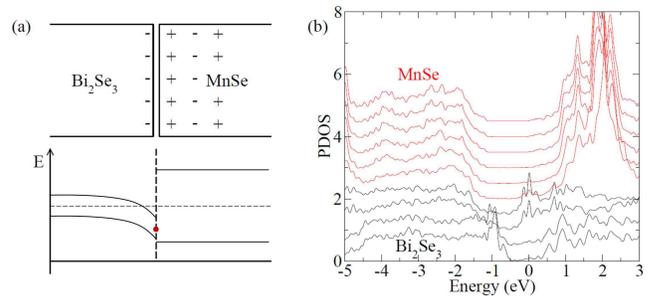}
\caption{(color online). (a) Illustration of the charge states at the
  Bi$_2$Se$_3$/MnSe interface, and a schematic picture of band
  bending, where the (red) dot at the interface represents the surface
  state Dirac point. (b) The calculated projected density-of-states of
  Bi$_2$Se$_3$ (black) and MnSe (red) for lattice sites across the
  surface, showing the band bending at the interface. Here each curve
  for MnSe represents one MnSe bilayer, and each curve for
  Bi$_2$Se$_3$ represents one BiSe bilayer in either the top or bottom
  half of the quintuple layers.}
\label{figure4}
\end{figure}

\noindent{\it Distance dependence of parameters $t$ and $M$.--} To
understand the behavior of large physical system, we investigate the
dependence of the effective magnetic field $M$ and inter-edge
interaction $t$ on the thickness of the MnSe slab in the superlattice
structure. First-principles band structure calculations of the 4-7
(composed of 4 QLs of Bi$_2$Se$_3$ and 7 layers of MnSe), 4-13, and
4-19 Bi$_2$Se$_3$/MnSe superlattices were fitted to the model solution
using equation~\ref{H_solution}. The fitting parameters $M$ and $t$
are plotted as a function of the MnSe layer number, as shown in
fig.~\ref{figure3}(b). The effective magnetic field $M$ does not vary
strongly as the number of MnSe layers increases, consistent with the
fact that it is mostly due to the magnetic exchange coupling between
the MnSe and the surface states of Bi$_2$Se$_3$ at the interface. In
contrast, the inter-edge interaction $t$ decreases rapidly as the
number of MnSe layers increases. This indicates that the inter-edge
interaction in smaller superlattices are mediated mostly through the
MnSe slab, instead of within the Bi$_2$Se$_3$ layers. The strong
effective magnetic field ($M \sim$ 27 meV) at this interface results
in a large gap opening (2$M \sim$ 54 meV) in realistic experimental
setup with thicker MnSe slab, which holds promise for magnetic
manipulation of TI surface states at room temperature.

In our calculation a perfect interface between MnSe and Bi$_2$Se$_3$
is assumed. In reality, the interface will probably contain impurities
and defects, so that the coupling between Mn and Se across the surface
is weaker. To see how the magnetic exchange coupling is affected by
the distance between Mn and Se atoms, we now keep fixed the atomic
coordinates within the Bi$_2$Se$_3$ and MnSe slabs, but varying the
separation $d$ between the Bi$_2$Se$_3$ and MnSe slabs, as shown in
fig.~\ref{figure3}(c). As the effective magnetic field $M$ is mostly
due to the exchange coupling between MnSe and Bi$_2$Se$_3$ across the
interface, we expect a strong dependence of $M$ on the separation
distance $d$. For the 4-13 Bi$_2$Se$_3$/MnSe superlattice, the
effective magnetic field $M$ is plotted in fig.~\ref{figure3}(d),
showing a strong dependence of $M$ as a function of the separation $d$
between Bi$_2$Se$_3$ and MnSe. This result indicates that the magnetic
exchange coupling at the TI-MI interface could be tuned by applying
external pressure.

\noindent{\bf Surface charge and band bending}

Although the Dirac cone is in the bulk gap of MnSe, from the band
structure shown in Fig. \ref{figure2} we see some additional hole-type
bands coexisting with the massive Dirac cone. These bands can be
understood as a consequence of the band bending occuring at the
heterostructure. Band bending at a heterojunction is common. In MnSe,
Mn$^{2+}$ and Se$^{2-}$ ions carry positive and negative charge,
respectively. Consequently, the Mn terminated surface considered here
carries a finite positive charge density. Since the Bi$_2$Se$_3$ is a
covalent material and has no surface charge, the net charge of the
heterojunction between these two materials is positive, leading to a
trap for the surface electrons. Therefore the bands bend down around
the interface, as is illutrated in Fig. \ref{figure4}(a). leading to
the additional surface state bands.

We studied the band bending effect by calculating the electronic
density-of-states projected to individual atomic planes. The projected
density-of-states (PDOS) of the Mn-Se atomic planes are shown in red
curves in fig.~\ref{figure4}(b), while the PDOS for the top and bottom
Bi-Se atomic planes in the Bi$_2$Se$_3$ slab are shown in black
curves. It is clear that very little band bending exists in MnSe,
while there is significant shift between the states of the first and
second QLs in Bi$_2$Se$_3$. The electronic states in the first QL of
Bi$_2$Se$_3$ is shifted down in energy compared to the second QL,
consistent with the direction of the interface electric dipole moment
shown in fig.~\ref{figure4}(a).

The additional surface states lead to difficulty in opening a full gap
at the surface, the solution of which is the main open question we
leave for future theoretical and experimental works. Possible
solutions include screening the surface charge by a top gate, or by
proper chemical doping at the junction.

%


\noindent{\bf Acknowledgements}

We acknowledge helpful discussions with Haijun Zhang and Shou-Cheng
Zhang. This work is supported by the Defense Advanced Research
Projects Agency Microsystems Technology Office, MesoDynamic
Architecture Program (MESO) through the contract number
N66001-11-1-4105. Computations were performed at the National Energy
Research Scientific Computing Center (NERSC), which is supported by
the Office of Science of the U.S. Department of Energy under Contract
No. DE-AC02-05CH11231.

\bibliography{TI,TI-MI}

\begin{thebibliography}{41}
\expandafter\ifx\csname natexlab\endcsname\relax\def\natexlab#1{#1}\fi
\expandafter\ifx\csname bibnamefont\endcsname\relax
  \def\bibnamefont#1{#1}\fi
\expandafter\ifx\csname bibfnamefont\endcsname\relax
  \def\bibfnamefont#1{#1}\fi
\expandafter\ifx\csname citenamefont\endcsname\relax
  \def\citenamefont#1{#1}\fi
\expandafter\ifx\csname url\endcsname\relax
  \def\url#1{\texttt{#1}}\fi
\expandafter\ifx\csname urlprefix\endcsname\relax\def\urlprefix{URL }\fi
\providecommand{\bibinfo}[2]{#2}
\providecommand{\eprint}[2][]{\url{#2}}

\bibitem[{\citenamefont{Moore}(2010)}]{moore2010}
\bibinfo{author}{\bibfnamefont{J.~E.} \bibnamefont{Moore}},
  \bibinfo{journal}{Nature} \textbf{\bibinfo{volume}{464}},
  \bibinfo{pages}{194} (\bibinfo{year}{2010}).

\bibitem[{\citenamefont{Hasan and Kane}(2010)}]{hasan2010}
\bibinfo{author}{\bibfnamefont{M.~Z.} \bibnamefont{Hasan}} \bibnamefont{and}
  \bibinfo{author}{\bibfnamefont{C.~L.} \bibnamefont{Kane}},
  \bibinfo{journal}{Rev. Mod. Phys.} \textbf{\bibinfo{volume}{82}},
  \bibinfo{pages}{3045} (\bibinfo{year}{2010}).

\bibitem[{\citenamefont{Qi and Zhang}(2011)}]{qi2011rmp}
\bibinfo{author}{\bibfnamefont{X.-L.} \bibnamefont{Qi}} \bibnamefont{and}
  \bibinfo{author}{\bibfnamefont{S.-C.} \bibnamefont{Zhang}},
  \bibinfo{journal}{Rev. Mod. Phys.} \textbf{\bibinfo{volume}{83}},
  \bibinfo{pages}{1057} (\bibinfo{year}{2011}).

\bibitem[{\citenamefont{Qi et~al.}(2008)\citenamefont{Qi, Hughes, and
  Zhang}}]{qi2008b}
\bibinfo{author}{\bibfnamefont{X.-L.} \bibnamefont{Qi}},
  \bibinfo{author}{\bibfnamefont{T.}~\bibnamefont{Hughes}}, \bibnamefont{and}
  \bibinfo{author}{\bibfnamefont{S.-C.} \bibnamefont{Zhang}},
  \bibinfo{journal}{Phys. Rev. B} \textbf{\bibinfo{volume}{78}},
  \bibinfo{pages}{195424} (\bibinfo{year}{2008}).

\bibitem[{\citenamefont{Essin et~al.}(2009)\citenamefont{Essin, Moore, and
  Vanderbilt}}]{essin2009}
\bibinfo{author}{\bibfnamefont{A.~M.} \bibnamefont{Essin}},
  \bibinfo{author}{\bibfnamefont{J.~E.} \bibnamefont{Moore}}, \bibnamefont{and}
  \bibinfo{author}{\bibfnamefont{D.}~\bibnamefont{Vanderbilt}},
  \bibinfo{journal}{Phys. Rev. Lett.} \textbf{\bibinfo{volume}{102}},
  \bibinfo{pages}{146805} (\bibinfo{year}{2009}).

\bibitem[{\citenamefont{Karch}(2009)}]{karch2009}
\bibinfo{author}{\bibfnamefont{A.}~\bibnamefont{Karch}},
  \bibinfo{journal}{Phys. Rev. Lett.} \textbf{\bibinfo{volume}{103}},
  \bibinfo{pages}{171601} (\bibinfo{year}{2009}).

\bibitem[{\citenamefont{Maciejko et~al.}(2010)\citenamefont{Maciejko, Qi, Drew,
  and Zhang}}]{maciejko2010}
\bibinfo{author}{\bibfnamefont{J.}~\bibnamefont{Maciejko}},
  \bibinfo{author}{\bibfnamefont{X.-L.} \bibnamefont{Qi}},
  \bibinfo{author}{\bibfnamefont{H.~D.} \bibnamefont{Drew}}, \bibnamefont{and}
  \bibinfo{author}{\bibfnamefont{S.-C.} \bibnamefont{Zhang}},
  \bibinfo{journal}{Phys. Rev. Lett.} \textbf{\bibinfo{volume}{105}},
  \bibinfo{pages}{166803} (\bibinfo{year}{2010}).

\bibitem[{\citenamefont{Wang-Kong~Tse}(2010)}]{tse2010}
\bibinfo{author}{\bibfnamefont{A.~H.~M.} \bibnamefont{Wang-Kong~Tse}},
  \bibinfo{journal}{Phys. Rev. Lett.} \textbf{\bibinfo{volume}{105}},
  \bibinfo{pages}{057401} (\bibinfo{year}{2010}).

\bibitem[{\citenamefont{Qi et~al.}(2009)\citenamefont{Qi, Li, Zang, and
  Zhang}}]{qi2009}
\bibinfo{author}{\bibfnamefont{X.-L.} \bibnamefont{Qi}},
  \bibinfo{author}{\bibfnamefont{R.}~\bibnamefont{Li}},
  \bibinfo{author}{\bibfnamefont{J.}~\bibnamefont{Zang}}, \bibnamefont{and}
  \bibinfo{author}{\bibfnamefont{S.-C.} \bibnamefont{Zhang}},
  \bibinfo{journal}{Science} \textbf{\bibinfo{volume}{323}},
  \bibinfo{pages}{1184} (\bibinfo{year}{2009}).

\bibitem[{\citenamefont{Rosenberg and Franz}(2010)}]{franz2010b}
\bibinfo{author}{\bibfnamefont{G.}~\bibnamefont{Rosenberg}} \bibnamefont{and}
  \bibinfo{author}{\bibfnamefont{M.}~\bibnamefont{Franz}},
  \bibinfo{journal}{Phys. Rev. B} \textbf{\bibinfo{volume}{82}},
  \bibinfo{pages}{035105} (\bibinfo{year}{2010}).

\bibitem[{\citenamefont{Rosenberg et~al.}(2010)\citenamefont{Rosenberg, Guo,
  and Franz}}]{franz2010c}
\bibinfo{author}{\bibfnamefont{G.}~\bibnamefont{Rosenberg}},
  \bibinfo{author}{\bibfnamefont{H.-M.} \bibnamefont{Guo}}, \bibnamefont{and}
  \bibinfo{author}{\bibfnamefont{M.}~\bibnamefont{Franz}},
  \bibinfo{journal}{Phys. Rev. B} \textbf{\bibinfo{volume}{82}},
  \bibinfo{pages}{041104} (\bibinfo{year}{2010}).

\bibitem[{\citenamefont{Yokoyama et~al.}(2010)\citenamefont{Yokoyama, Zang, and
  Nagaosa}}]{yokoyama2010b}
\bibinfo{author}{\bibfnamefont{T.}~\bibnamefont{Yokoyama}},
  \bibinfo{author}{\bibfnamefont{J.}~\bibnamefont{Zang}}, \bibnamefont{and}
  \bibinfo{author}{\bibfnamefont{N.}~\bibnamefont{Nagaosa}},
  \bibinfo{journal}{Phys. Rev. B} \textbf{\bibinfo{volume}{81}},
  \bibinfo{pages}{241410} (\bibinfo{year}{2010}).

\bibitem[{\citenamefont{Nomura and Nagaosa}(2010)}]{nomura2010}
\bibinfo{author}{\bibfnamefont{K.}~\bibnamefont{Nomura}} \bibnamefont{and}
  \bibinfo{author}{\bibfnamefont{N.}~\bibnamefont{Nagaosa}},
  \bibinfo{journal}{Phys. Rev. B} \textbf{\bibinfo{volume}{82}},
  \bibinfo{pages}{161401(R)} (\bibinfo{year}{2010}).

\bibitem[{\citenamefont{Aguilar et~al.}(2011)\citenamefont{Aguilar, Stier, Liu,
  Bilbro, George, Bansal, Cerne, Markelz, Oh, and Armitage}}]{aguilar2011}
\bibinfo{author}{\bibfnamefont{R.~V.} \bibnamefont{Aguilar}},
  \bibinfo{author}{\bibfnamefont{A.}~\bibnamefont{Stier}},
  \bibinfo{author}{\bibfnamefont{W.}~\bibnamefont{Liu}},
  \bibinfo{author}{\bibfnamefont{L.}~\bibnamefont{Bilbro}},
  \bibinfo{author}{\bibfnamefont{D.}~\bibnamefont{George}},
  \bibinfo{author}{\bibfnamefont{N.}~\bibnamefont{Bansal}},
  \bibinfo{author}{\bibfnamefont{J.}~\bibnamefont{Cerne}},
  \bibinfo{author}{\bibfnamefont{A.}~\bibnamefont{Markelz}},
  \bibinfo{author}{\bibfnamefont{S.}~\bibnamefont{Oh}}, \bibnamefont{and}
  \bibinfo{author}{\bibfnamefont{N.}~\bibnamefont{Armitage}},
  \emph{\bibinfo{title}{Thz response and colossal magneto-electric effect in
  the topological insulator bi$_2$se$_3$}}, \bibinfo{howpublished}{e-print
  arXiv:1105.0237} (\bibinfo{year}{2011}).

\bibitem[{\citenamefont{{LaForge} et~al.}(2010)\citenamefont{{LaForge},
  Frenzel, Pursley, Lin, Liu, Shi, and Basov}}]{laforge2010}
\bibinfo{author}{\bibfnamefont{A.~D.} \bibnamefont{{LaForge}}},
  \bibinfo{author}{\bibfnamefont{A.}~\bibnamefont{Frenzel}},
  \bibinfo{author}{\bibfnamefont{B.~C.} \bibnamefont{Pursley}},
  \bibinfo{author}{\bibfnamefont{T.}~\bibnamefont{Lin}},
  \bibinfo{author}{\bibfnamefont{X.}~\bibnamefont{Liu}},
  \bibinfo{author}{\bibfnamefont{J.}~\bibnamefont{Shi}}, \bibnamefont{and}
  \bibinfo{author}{\bibfnamefont{D.~N.} \bibnamefont{Basov}},
  \bibinfo{journal}{Phys. Rev. B} \textbf{\bibinfo{volume}{81}},
  \bibinfo{pages}{125120} (\bibinfo{year}{2010}).

\bibitem[{\citenamefont{Sushkov et~al.}(2010)\citenamefont{Sushkov, Jenkins,
  Schmadel, Butch, Paglione, and Drew}}]{sushkov2010}
\bibinfo{author}{\bibfnamefont{A.~B.} \bibnamefont{Sushkov}},
  \bibinfo{author}{\bibfnamefont{G.~S.} \bibnamefont{Jenkins}},
  \bibinfo{author}{\bibfnamefont{D.~C.} \bibnamefont{Schmadel}},
  \bibinfo{author}{\bibfnamefont{N.~P.} \bibnamefont{Butch}},
  \bibinfo{author}{\bibfnamefont{J.}~\bibnamefont{Paglione}}, \bibnamefont{and}
  \bibinfo{author}{\bibfnamefont{H.~D.} \bibnamefont{Drew}},
  \bibinfo{journal}{Phys. Rev. B} \textbf{\bibinfo{volume}{82}},
  \bibinfo{pages}{125110} (\bibinfo{year}{2010}).

\bibitem[{\citenamefont{Jenkins et~al.}(2010)\citenamefont{Jenkins, Sushkov,
  Schmadel, Butch, Syers, Paglione, and Drew}}]{jenkins2010}
\bibinfo{author}{\bibfnamefont{G.~S.} \bibnamefont{Jenkins}},
  \bibinfo{author}{\bibfnamefont{A.~B.} \bibnamefont{Sushkov}},
  \bibinfo{author}{\bibfnamefont{D.~C.} \bibnamefont{Schmadel}},
  \bibinfo{author}{\bibfnamefont{N.~P.} \bibnamefont{Butch}},
  \bibinfo{author}{\bibfnamefont{P.}~\bibnamefont{Syers}},
  \bibinfo{author}{\bibfnamefont{J.}~\bibnamefont{Paglione}}, \bibnamefont{and}
  \bibinfo{author}{\bibfnamefont{H.~D.} \bibnamefont{Drew}},
  \bibinfo{journal}{Phys. Rev. B} \textbf{\bibinfo{volume}{82}},
  \bibinfo{pages}{125120} (\bibinfo{year}{2010}).

\bibitem[{\citenamefont{Liu et~al.}(2009)\citenamefont{Liu, Liu, Xu, Qi, and
  Zhang}}]{liu2009}
\bibinfo{author}{\bibfnamefont{Q.}~\bibnamefont{Liu}},
  \bibinfo{author}{\bibfnamefont{C.-X.} \bibnamefont{Liu}},
  \bibinfo{author}{\bibfnamefont{C.}~\bibnamefont{Xu}},
  \bibinfo{author}{\bibfnamefont{X.-L.} \bibnamefont{Qi}}, \bibnamefont{and}
  \bibinfo{author}{\bibfnamefont{S.-C.} \bibnamefont{Zhang}},
  \bibinfo{journal}{Phys. Rev. Lett.} \textbf{\bibinfo{volume}{102}},
  \bibinfo{pages}{156603} (\bibinfo{year}{2009}).

\bibitem[{\citenamefont{Yu et~al.}(2010)\citenamefont{Yu, Zhang, Zhang, Zhang,
  Dai, and Fang}}]{yu2010}
\bibinfo{author}{\bibfnamefont{R.}~\bibnamefont{Yu}},
  \bibinfo{author}{\bibfnamefont{W.}~\bibnamefont{Zhang}},
  \bibinfo{author}{\bibfnamefont{H.~J.} \bibnamefont{Zhang}},
  \bibinfo{author}{\bibfnamefont{S.~C.} \bibnamefont{Zhang}},
  \bibinfo{author}{\bibfnamefont{X.}~\bibnamefont{Dai}}, \bibnamefont{and}
  \bibinfo{author}{\bibfnamefont{Z.}~\bibnamefont{Fang}},
  \bibinfo{journal}{Science} \textbf{\bibinfo{volume}{329}},
  \bibinfo{pages}{61} (\bibinfo{year}{2010}).

\bibitem[{\citenamefont{Chen et~al.}(2010)\citenamefont{Chen, Chu, Analytis,
  Liu, Igarashi, Kuo, Qi, Mo, Moore, Lu et~al.}}]{chen2010b}
\bibinfo{author}{\bibfnamefont{Y.~L.} \bibnamefont{Chen}},
  \bibinfo{author}{\bibfnamefont{J.-H.} \bibnamefont{Chu}},
  \bibinfo{author}{\bibfnamefont{J.~G.} \bibnamefont{Analytis}},
  \bibinfo{author}{\bibfnamefont{Z.~K.} \bibnamefont{Liu}},
  \bibinfo{author}{\bibfnamefont{K.}~\bibnamefont{Igarashi}},
  \bibinfo{author}{\bibfnamefont{H.-H.} \bibnamefont{Kuo}},
  \bibinfo{author}{\bibfnamefont{X.~L.} \bibnamefont{Qi}},
  \bibinfo{author}{\bibfnamefont{S.~K.} \bibnamefont{Mo}},
  \bibinfo{author}{\bibfnamefont{R.~G.} \bibnamefont{Moore}},
  \bibinfo{author}{\bibfnamefont{D.~H.} \bibnamefont{Lu}},
  \bibnamefont{et~al.}, \bibinfo{journal}{Science}
  \textbf{\bibinfo{volume}{329}}, \bibinfo{pages}{659} (\bibinfo{year}{2010}).

\bibitem[{\citenamefont{Xu et~al.}()\citenamefont{Xu, Neupane, Liu, Zhang,
  Richardella, Wray, Alidoust, Leandersson, Balasubramanian, Sachez-Barriga
  et~al.}}]{hasan2012}
\bibinfo{author}{\bibfnamefont{S.-Y.} \bibnamefont{Xu}},
  \bibinfo{author}{\bibfnamefont{M.}~\bibnamefont{Neupane}},
  \bibinfo{author}{\bibfnamefont{C.}~\bibnamefont{Liu}},
  \bibinfo{author}{\bibfnamefont{D.~M.} \bibnamefont{Zhang}},
  \bibinfo{author}{\bibfnamefont{A.}~\bibnamefont{Richardella}},
  \bibinfo{author}{\bibfnamefont{L.~A.} \bibnamefont{Wray}},
  \bibinfo{author}{\bibfnamefont{N.}~\bibnamefont{Alidoust}},
  \bibinfo{author}{\bibfnamefont{M.}~\bibnamefont{Leandersson}},
  \bibinfo{author}{\bibfnamefont{T.}~\bibnamefont{Balasubramanian}},
  \bibinfo{author}{\bibfnamefont{J.}~\bibnamefont{Sachez-Barriga}},
  \bibnamefont{et~al.}, \emph{\bibinfo{title}{Magnetically induced spin
  reorientation and magnetic transition on the surface of a topological
  insulator}}, \bibinfo{howpublished}{e-print arXiv:1206.2090}.

\bibitem[{\citenamefont{Liu et~al.}(2012)\citenamefont{Liu, Zhang, Chang,
  Zhang, Feng, Li, He, Wang, Chen, Dai et~al.}}]{liu2012}
\bibinfo{author}{\bibfnamefont{M.}~\bibnamefont{Liu}},
  \bibinfo{author}{\bibfnamefont{J.}~\bibnamefont{Zhang}},
  \bibinfo{author}{\bibfnamefont{C.-Z.} \bibnamefont{Chang}},
  \bibinfo{author}{\bibfnamefont{Z.}~\bibnamefont{Zhang}},
  \bibinfo{author}{\bibfnamefont{X.}~\bibnamefont{Feng}},
  \bibinfo{author}{\bibfnamefont{K.}~\bibnamefont{Li}},
  \bibinfo{author}{\bibfnamefont{K.}~\bibnamefont{He}},
  \bibinfo{author}{\bibfnamefont{L.-l.} \bibnamefont{Wang}},
  \bibinfo{author}{\bibfnamefont{X.}~\bibnamefont{Chen}},
  \bibinfo{author}{\bibfnamefont{X.}~\bibnamefont{Dai}}, \bibnamefont{et~al.},
  \bibinfo{journal}{Phys. Rev. Lett.} \textbf{\bibinfo{volume}{108}},
  \bibinfo{pages}{036805} (\bibinfo{year}{2012}),
  \urlprefix\url{http://link.aps.org/doi/10.1103/PhysRevLett.108.036805}.

\bibitem[{\citenamefont{Chang et~al.}(2011)\citenamefont{Chang, Zhang, Liu,
  Zhang, Feng, Li, Wang, Chen, Dai, Fang et~al.}}]{chang2011}
\bibinfo{author}{\bibfnamefont{C.-Z.} \bibnamefont{Chang}},
  \bibinfo{author}{\bibfnamefont{J.-S.} \bibnamefont{Zhang}},
  \bibinfo{author}{\bibfnamefont{M.-H.} \bibnamefont{Liu}},
  \bibinfo{author}{\bibfnamefont{Z.-C.} \bibnamefont{Zhang}},
  \bibinfo{author}{\bibfnamefont{X.}~\bibnamefont{Feng}},
  \bibinfo{author}{\bibfnamefont{K.}~\bibnamefont{Li}},
  \bibinfo{author}{\bibfnamefont{L.-L.} \bibnamefont{Wang}},
  \bibinfo{author}{\bibfnamefont{X.}~\bibnamefont{Chen}},
  \bibinfo{author}{\bibfnamefont{X.}~\bibnamefont{Dai}},
  \bibinfo{author}{\bibfnamefont{Z.}~\bibnamefont{Fang}}, \bibnamefont{et~al.},
  \emph{\bibinfo{title}{Carrier-independent ferromagnetism and giant anomalous
  hall effect in magnetic topological insulator}},
  \bibinfo{howpublished}{e-print arXiv:1108.4754} (\bibinfo{year}{2011}).

\bibitem[{\citenamefont{P\'erez~Vicente
  et~al.}(1999)\citenamefont{P\'erez~Vicente, Tirado, Adouby, Jumas,
  Abba~Tour\'e, and Kra}}]{Bi2Se3_struc}
\bibinfo{author}{\bibfnamefont{C.}~\bibnamefont{P\'erez~Vicente}},
  \bibinfo{author}{\bibfnamefont{J.~L.} \bibnamefont{Tirado}},
  \bibinfo{author}{\bibfnamefont{K.}~\bibnamefont{Adouby}},
  \bibinfo{author}{\bibfnamefont{J.~C.} \bibnamefont{Jumas}},
  \bibinfo{author}{\bibfnamefont{A.}~\bibnamefont{Abba~Tour\'e}},
  \bibnamefont{and} \bibinfo{author}{\bibfnamefont{G.}~\bibnamefont{Kra}},
  \bibinfo{journal}{Inorg.\ Chem.} \textbf{\bibinfo{volume}{38}},
  \bibinfo{pages}{2131} (\bibinfo{year}{1999}).

\bibitem[{\citenamefont{Feutelais et~al.}(1993)\citenamefont{Feutelais,
  Legendre, Rodier, and Agafonov}}]{Bi2Te3_struc}
\bibinfo{author}{\bibfnamefont{Y.}~\bibnamefont{Feutelais}},
  \bibinfo{author}{\bibfnamefont{B.}~\bibnamefont{Legendre}},
  \bibinfo{author}{\bibfnamefont{N.}~\bibnamefont{Rodier}}, \bibnamefont{and}
  \bibinfo{author}{\bibfnamefont{V.}~\bibnamefont{Agafonov}},
  \bibinfo{journal}{Mater.\ Res.\ Bull.} \textbf{\bibinfo{volume}{28}},
  \bibinfo{pages}{591} (\bibinfo{year}{1993}).

\bibitem[{\citenamefont{Anderson and Krause}(1974)}]{Sb2Te3_struc}
\bibinfo{author}{\bibfnamefont{T.~L.} \bibnamefont{Anderson}} \bibnamefont{and}
  \bibinfo{author}{\bibfnamefont{H.~B.} \bibnamefont{Krause}},
  \bibinfo{journal}{Acta Cryst.} \textbf{\bibinfo{volume}{B30}},
  \bibinfo{pages}{1307} (\bibinfo{year}{1974}).

\bibitem[{\citenamefont{McWhan et~al.}(1966)\citenamefont{McWhan, Souers, and
  Jura}}]{EuO_struc}
\bibinfo{author}{\bibfnamefont{D.~B.} \bibnamefont{McWhan}},
  \bibinfo{author}{\bibfnamefont{P.~C.} \bibnamefont{Souers}},
  \bibnamefont{and} \bibinfo{author}{\bibfnamefont{G.}~\bibnamefont{Jura}},
  \bibinfo{journal}{Phys. Rev.} \textbf{\bibinfo{volume}{143}},
  \bibinfo{pages}{385} (\bibinfo{year}{1966}).

\bibitem[{\citenamefont{Palazii and Bretey}(1989)}]{EuS_struc}
\bibinfo{author}{\bibfnamefont{M.}~\bibnamefont{Palazii}} \bibnamefont{and}
  \bibinfo{author}{\bibfnamefont{E.}~\bibnamefont{Bretey}},
  \bibinfo{journal}{Mater.\ Res.\ Bull.} \textbf{\bibinfo{volume}{24}},
  \bibinfo{pages}{695} (\bibinfo{year}{1989}).

\bibitem[{\citenamefont{Westerholt and Bach}(1985)}]{EuSe_struc}
\bibinfo{author}{\bibfnamefont{K.}~\bibnamefont{Westerholt}} \bibnamefont{and}
  \bibinfo{author}{\bibfnamefont{H.}~\bibnamefont{Bach}},
  \bibinfo{journal}{Phys.\ Rev.\ B} \textbf{\bibinfo{volume}{31}},
  \bibinfo{pages}{7151} (\bibinfo{year}{1985}).

\bibitem[{\citenamefont{Jacobson and Fender}(1970)}]{MnSe_struc}
\bibinfo{author}{\bibfnamefont{A.~J.} \bibnamefont{Jacobson}} \bibnamefont{and}
  \bibinfo{author}{\bibfnamefont{B.~E.~F.} \bibnamefont{Fender}},
  \bibinfo{journal}{J.\ Chem.\ Phys.} \textbf{\bibinfo{volume}{52}},
  \bibinfo{pages}{4563} (\bibinfo{year}{1970}).

\bibitem[{\citenamefont{Noor}(1987)}]{MnTe_struc}
\bibinfo{author}{\bibfnamefont{S.~S.~A.} \bibnamefont{Noor}},
  \bibinfo{journal}{J.\ Appl.\ Phys.} \textbf{\bibinfo{volume}{61}},
  \bibinfo{pages}{3549} (\bibinfo{year}{1987}).

\bibitem[{\citenamefont{Ali and Felimban}(1989)}]{RbMnCl3_struc}
\bibinfo{author}{\bibfnamefont{E.~M.} \bibnamefont{Ali}} \bibnamefont{and}
  \bibinfo{author}{\bibfnamefont{A.~A.} \bibnamefont{Felimban}},
  \bibinfo{journal}{Aust.\ J.\ Phys.} \textbf{\bibinfo{volume}{42}},
  \bibinfo{pages}{307} (\bibinfo{year}{1989}).

\bibitem[{\citenamefont{Perdew et~al.}(1996)\citenamefont{Perdew, Burke, and
  Ernzerhof}}]{PBE1996}
\bibinfo{author}{\bibfnamefont{J.~P.} \bibnamefont{Perdew}},
  \bibinfo{author}{\bibfnamefont{K.}~\bibnamefont{Burke}}, \bibnamefont{and}
  \bibinfo{author}{\bibfnamefont{M.}~\bibnamefont{Ernzerhof}},
  \bibinfo{journal}{Phys.\ Rev.\ Lett.} \textbf{\bibinfo{volume}{77}},
  \bibinfo{pages}{3865} (\bibinfo{year}{1996}).

\bibitem[{\citenamefont{Bl\"ochl}(1994)}]{Blochl1994}
\bibinfo{author}{\bibfnamefont{P.~E.} \bibnamefont{Bl\"ochl}},
  \bibinfo{journal}{Phys.\ Rev.\ B} \textbf{\bibinfo{volume}{50}},
  \bibinfo{pages}{17953} (\bibinfo{year}{1994}).

\bibitem[{\citenamefont{Kresse and Furthm\"uller}(1996)}]{Kresse1996}
\bibinfo{author}{\bibfnamefont{G.}~\bibnamefont{Kresse}} \bibnamefont{and}
  \bibinfo{author}{\bibfnamefont{J.}~\bibnamefont{Furthm\"uller}},
  \bibinfo{journal}{Phys.\ Rev.\ B} \textbf{\bibinfo{volume}{54}},
  \bibinfo{pages}{11169} (\bibinfo{year}{1996}).

\bibitem[{\citenamefont{Kresse and Joubert}(1999)}]{Kresse1999}
\bibinfo{author}{\bibfnamefont{G.}~\bibnamefont{Kresse}} \bibnamefont{and}
  \bibinfo{author}{\bibfnamefont{D.}~\bibnamefont{Joubert}},
  \bibinfo{journal}{Phys.\ Rev.\ B} \textbf{\bibinfo{volume}{59}},
  \bibinfo{pages}{1758} (\bibinfo{year}{1999}).

\bibitem[{\citenamefont{Anisimov et~al.}(1991)\citenamefont{Anisimov, Zaanen,
  and Andersen}}]{Anisimov1991}
\bibinfo{author}{\bibfnamefont{V.~I.} \bibnamefont{Anisimov}},
  \bibinfo{author}{\bibfnamefont{J.}~\bibnamefont{Zaanen}}, \bibnamefont{and}
  \bibinfo{author}{\bibfnamefont{O.~K.} \bibnamefont{Andersen}},
  \bibinfo{journal}{Phys.\ Rev.\ B} \textbf{\bibinfo{volume}{44}},
  \bibinfo{pages}{943} (\bibinfo{year}{1991}).

\bibitem[{\citenamefont{Liechtenstein et~al.}(1995)\citenamefont{Liechtenstein,
  Anisimov, and Zaanen}}]{Liechtenstein1995}
\bibinfo{author}{\bibfnamefont{A.~I.} \bibnamefont{Liechtenstein}},
  \bibinfo{author}{\bibfnamefont{V.~I.} \bibnamefont{Anisimov}},
  \bibnamefont{and} \bibinfo{author}{\bibfnamefont{J.}~\bibnamefont{Zaanen}},
  \bibinfo{journal}{Phys.\ Rev.\ B} \textbf{\bibinfo{volume}{52}},
  \bibinfo{pages}{R5467} (\bibinfo{year}{1995}).

\bibitem[{\citenamefont{Youn}(2005)}]{Youn2005}
\bibinfo{author}{\bibfnamefont{S.~J.} \bibnamefont{Youn}},
  \bibinfo{journal}{Journal of Magnetics} \textbf{\bibinfo{volume}{10}},
  \bibinfo{pages}{71} (\bibinfo{year}{2005}).

\bibitem[{\citenamefont{Amiri et~al.}(2011)\citenamefont{Amiri, Hashemifar, and
  Akbarzadeh}}]{Amiri2011}
\bibinfo{author}{\bibfnamefont{P.}~\bibnamefont{Amiri}},
  \bibinfo{author}{\bibfnamefont{S.~J.} \bibnamefont{Hashemifar}},
  \bibnamefont{and}
  \bibinfo{author}{\bibfnamefont{H.}~\bibnamefont{Akbarzadeh}},
  \bibinfo{journal}{Phys.\ Rev.\ B} \textbf{\bibinfo{volume}{83}},
  \bibinfo{pages}{165424} (\bibinfo{year}{2011}).

\bibitem[{\citenamefont{Liu et~al.}(2010)\citenamefont{Liu, Qi, Zhang, Dai,
  Fang, and Zhang}}]{liu2010}
\bibinfo{author}{\bibfnamefont{C.-X.} \bibnamefont{Liu}},
  \bibinfo{author}{\bibfnamefont{X.-L.} \bibnamefont{Qi}},
  \bibinfo{author}{\bibfnamefont{H.}~\bibnamefont{Zhang}},
  \bibinfo{author}{\bibfnamefont{X.}~\bibnamefont{Dai}},
  \bibinfo{author}{\bibfnamefont{Z.}~\bibnamefont{Fang}}, \bibnamefont{and}
  \bibinfo{author}{\bibfnamefont{S.-C.} \bibnamefont{Zhang}},
  \bibinfo{journal}{Phys. Rev. B} \textbf{\bibinfo{volume}{82}},
  \bibinfo{pages}{045122} (\bibinfo{year}{2010}).

\end{thebibliography}

\newpage

\begin{widetext}

\appendix

\section{In-plane magnetization}
As a comparison, we also calculated the bandstructure of
Bi$_2$Se$_3$/MnSe superlattice with Mn spins along the in-plane [100]
direction. The bandstructure along $k_y$ ($\Gamma$-K) direction with
projection of the bands to the Bi and Se atoms of the top and bottom
surfaces is shown in fig.~\ref{figure-S1}(a). Similar to the case of
[001] spin orietation shown in Fig.~\ref{figure2}, the Dirac-cone
states are located about 0.4 eV below the bulk gap of
Bi$_2$Se$_3$. One prominent feature is that the Dirac-cone states of
the top and bottom surfaces are shifted to opposite directions along
$k_y$ axis. For a Fermi energy above the Dirac cone, the Fermi
surfaces from the top and bottom surfaces of Bi$_2$Se$_3$ will shift
to opposite directions in the momentum space. Fig.~\ref{figure-S1}(b)
illustrates the opposite shiftof the two Fermi surfaces along $k_y$
direction, and the corresponding spin directions.

By fitting the effective model (\ref{effective_H}) with an in-plane
exchange field, we have extracted the exchange coupling $M$ and
inter-edge interaction $t$. For the 4-13 superlattice structure with
in-plane magnetization, we obtain $M$ = 9 meV, and $t$ = 19 meV. The
value of $t$ mostly depends on the thickness of MnSe and Bi$_2$Se$_3$
slabs, so $t$ in the case of in-plane magnetization is similar to the
value for perpendicular magnetization. However the exchange coupling
is quite anisotropic, and it is much smaller in the case of in-plane
magnetization compared to the perpendicular one.


\section{${\rm Sb_2Te_3/MnTe}$ interface}
Other materials in our study include the Sb$_2$Te$_3$/MnTe
heterostructure.  The calculated bandstructure of the 4-13
superlattice with projection of the bands to the Sb and Te atoms at
the top and bottom surfaces is shown in fig.~\ref{figure-S2}(a). In
contrast to the Bi$_2$Se$_3$/MnSe heterostructure, there is
complicated hybridization of bulk and surface states in the
Sb$_2$Te$_3$/MnTe heterostructure, and the Dirac-cone feature can not
be easily identified.

We project the bands separately to the Sb$_2$Te$_3$ slab and the MnTe
slab in the superlattice, as shown in
fig.~\ref{figure-S2}(b). Although the bulk band gap of MnTe at
$\Gamma$ point is much larger than Sb$_2$Te$_3$, the bulk band gap of
Sb$_2$Te$_3$ is located close in energy to the top of valence band of
MnTe, and the Dirac surface states overlap in energy with the valence
band of MnTe. Therefore it is less likely to realize a fully gapped
system in Sb$_2$Te$_3$/MnTe heterostructure.

\section{${\rm Bi_2Se_3/MnTe}$ interface}
We have also studied the Bi$_2$Se$_3$/MnTe heterostructure with
supercell composed of 4 QLs of Bi$_2$Se$_3$ and 7 layers of MnTe.  The
calculated bandstructure with projection of the bands to the Bi and Se
atoms at the top and bottom surfaces is shown in
fig.~\ref{figure-S3}. Similar to the Sb$_2$Te$_3$/MnTe
heterostructure, the surface states and bulk states in the
Bi$_2$Se$_3$/MnTe heterostructure are also stronly hybridized, and the
Dirac-cone feature can not be easily identified.

%
\begin{figure}[h]
\includegraphics[width=85mm]{fig-S1a.eps} \hspace{20mm}\includegraphics[width=25mm]{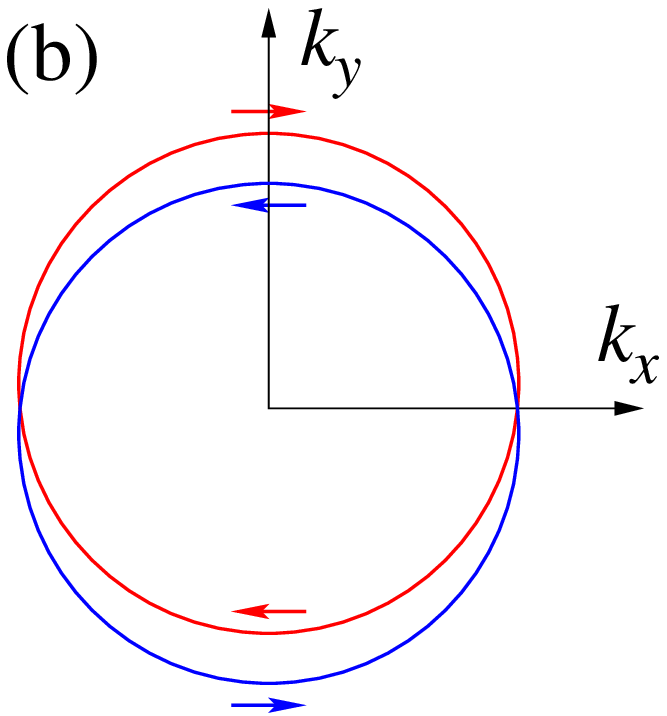}
\caption{(Color online) (a) Bandstructure of the Bi$_2$Se$_3$/MnSe
  superlattice along $k_y$ ($\Gamma$-K) direction. The red and blue
  symbols show the projection of the wavefunctions to the Bi and Se
  atoms at the top and bottom surfaces. (b) a schematic showing the
  Fermi surfaces of $n$-doped surface states at the two surfaces of
  Bi$_2$Se$_3$. The Fermi energy is taken at the dashed line in
  (a). The calculation is performed for superlattice structure
  composed of 4 QLs of Bi$_2$Se$_3$ and 13 layers of MnSe, with Mn
  spins along [100] direction.}
\label{figure-S1}
\end{figure}

%
\begin{figure}[h]
\includegraphics[width=85mm]{fig-S2a.eps}\includegraphics[width=85mm]{fig-S2b.eps}
\caption{(Color online) Bandstructure of the Sb$_2$Te$_3$/MnTe
  superlattice. The two panels show the projection of the
  wavefunctions to the Sb and Te atoms at the top (red symbols) and
  bottom (blue symbols) surfaces (a), and that to the two materials
  Sb$_2$Te$_3$ and MnTe (b). The calculation is done for the
  superlattice structure composed of 4 QLs of Sb$_2$Te$_3$ and 13
  layers of MnTe, with Mn spins along [001] direction.}
\label{figure-S2}
\end{figure}

%
\begin{figure}
\vspace{3mm}
\includegraphics[width=85mm]{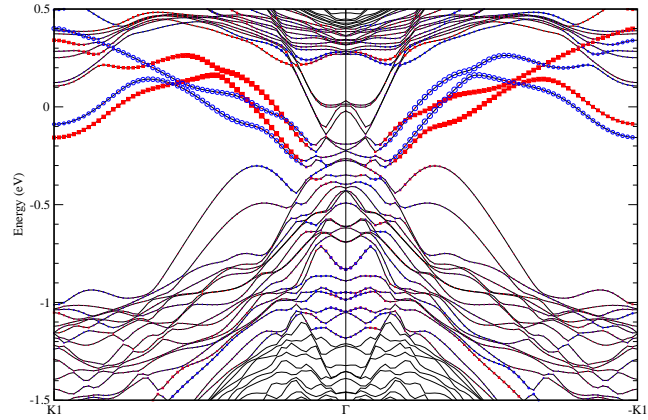}
\caption{(Color online) Bandstructure of the Bi$_2$Se$_3$/MnTe
  superlattice, showing the projection of the wavefunctions to the Bi
  and Se atoms at the top (red symbols) and bottom (blue symbols)
  surfaces. The calculation is done for the superlattice structure
  composed of 4 QLs of Bi$_2$Se$_3$ and 7 layers of MnTe, with Mn
  spins along [001] direction.}
\label{figure-S3}
\end{figure}

\end{widetext}

\end{document}